\documentclass[12pt]{iopart}

\usepackage{epsfig}
\usepackage{color}
\usepackage{graphics}
\usepackage{amssymb}
\usepackage{iopams}
\usepackage{mathrsfs}

\begin{document}

\title{Anomalous, non-Gaussian tracer diffusion in heterogeneously crowded
environments}

\author{Surya K. Ghosh$^{\dagger,\ddagger}$, Andrey G. Cherstvy$^\dagger$,
Denis S. Grebenkov$^\ddagger$, and Ralf Metzler$^{\dagger,\sharp,1}$}
\address{$\dagger$ Institute for Physics \& Astronomy, University of Potsdam,
14476 Potsdam-Golm, Germany\\
$\ddagger$ Physics of Condensed Matter, CNRS {\'E}cole Polytechnique, 91128
Palaiseau, France\\
$\sharp$ Department of Physics, Tampere University of Technology, 33101 Tampere,
Finland}
\ead{$^1$rmetzler@uni-potsdam.de}

\date{\today}

\begin{abstract}
A topic of intense current investigation pursues the question how the highly
crowded environment of biological cells affects the dynamic properties of
passively diffusing particles. Motivated by recent experiments we report
results of extensive simulations of the motion of a finite sized tracer
particle in a heterogeneously crowded environment. For given spatial distributions
of monodisperse crowders we demonstrate how anomalous diffusion with strongly
non-Gaussian features arises in this model system. We investigate both
biologically relevant situations of particles released either at the surface of an
inner domain (\emph{nucleus\/}), or at the outer boundary
(\emph{cell membrane\/}), exhibiting distinctly different behaviour of the
observed anomalous diffusion for heterogeneous crowder distributions.
Specifically we reveal an extremely asymmetric spreading of the tracer even at
moderate crowding fractions.
In addition to the standard mean squared displacement and the local diffusion
exponent of the tracer particles we investigate the magnitude and the amplitude
scatter of the time averaged mean squared displacement of individual trajectories,
the non-Gaussianity parameter, and the van Hove correlation function of the
particle displacements. We also quantify how the average tracer diffusivity
varies with the position in the domain with heterogeneous radial distribution
of the crowders and examine the behaviour of the survival probability and the
dynamics of first passage events of the tracer. Finally, we discuss the
relevance of our results to single particle tracking measurements in 
biological cells.
\end{abstract}

\section{Introduction} 

The cytoplasmic fluid of living cells is a supercrowded medium
\cite{golding}, in which biomacromolecules occupy volume fractions 
reaching $30\%$ and higher \cite{zimm91,zimm93,zhou04,mint08,elco10}. 
This macromolecular crowding (MMC) affects the diffusion of larger passive
molecules, endogenous as well as artificially introduced submicron tracer
particles, and cellular components
\cite{fran13}. One of the central observations is the existence of transient
but often very extended anomalous diffusion \cite{bouchaud,report} with the
sublinear scaling
\begin{equation}
\left<\textbf{r}^2(t)\right>\simeq K_{\beta}t^{\beta}
\label{eq-msd-sub}
\end{equation}
of the mean squared displacement (MSD) of the diffusing particles with the 
anomalous diffusion exponent in the subdiffusive range $0<\beta<1$
\cite{fran13,pt}. Here $K_\beta$ is the generalised diffusion coefficient with
units $\mathrm{cm}^2/\mathrm{sec}^{\beta}$. Subdiffusion in the crowded
cytoplasm of living cells was observed for fluorescently labelled and
autofluorescent small proteins \cite{frad05,gratton}, labelled polymeric dextrane
\cite{weis04} and messenger RNA \cite{golding,spak10}, chromosomal loci and
telomeres \cite{spak10,gari09}, as well as submicron endogenous granules
\cite{jeon11,bertseva,tabei} and viruses \cite{seis01}. Subdiffusion was also reported
for the motion of tracer particles in artificially crowded environments \cite{yves,
pan,weis09,weis12,lene1,greb13,greb14,weis14,hayward}, and it also occurs in cell membranes,
as shown experimentally \cite{weis03,krap11,lape15} and by molecular dynamics as
well as coarse grained simulations \cite{kneller,jae_membrane,grub14,akimoto}. 
In addition, active transport processes in living cells may lead to superdiffusion 
with $1<\beta<2$  \cite{elbaum,daphne,nguyen,christine}.

Particle diffusion in crowded and structured environments has been in the focus of
a number of computer simulations \cite{saxt96,fran06,netz_gel,godec,berr13,trov14}
and theoretical studies \cite{fran13,fran06,saxt14,rubinshteyn}. The observed
anomalous diffusion in such systems is addressed by various generalised stochastic
processes, as summarised in references \cite{soko12,bres13,metz14}. Specifically
in an environment of densely packed, freely moving crowders the tracer diffusion
was demonstrated to follow Brownian motion at sufficiently long times \cite{berr13},
whereas for crowders confined by a potential and for static crowders the tracer
diffusion features a very extended albeit ultimately transient subdiffusive regime
\cite{berr13,ghos15}.

The current study is motivated by recent experimental evidence of an inherently
polydisperse mixture of crowding proteins in both the bacterial and eukaryotic
cytoplasm \cite{elco10,berr13,trov14}. Moreover, the distribution of crowders in
the cell was shown to be rather heterogeneous, giving rise to a faster particle
mobility of small tracer proteins near the cell nucleus of surface-adhered
eukaryotic cells \cite{lang11}, see also the heterogeneous diffusivity map
in reference \cite{elf11}. These properties of the cell cytoplasm impose severe
restrictions on the rates of biochemical reactions \cite{mint01,foff10} including
those involving polymer dynamics \cite{shin15a,shin15b,fyl,ha} and often impair the
diffusion of particles inside cells \cite{berr13,trov14,ghos15}.

Here we address two aspects of crowding, a finite size of the tracer and a
heterogeneous
distribution of crowders in a two dimensional, circular model cell with a central
nucleus region. In the space between the outer cell boundary and the inner nucleus
we place either homogeneously or heterogeneously distributed monodisperse crowders, 
and then simulate the motion of a finite sized tracer particle through this static
crowder configuration. We investigate the two biologically relevant scenarios of in-out
(from nucleus to cell boundary) versus out-in  (from cell boundary
to nucleus) tracer diffusion, finding fundamental differences in the observed
dynamics. From extensive simulations we determine the particle distribution for
different crowding environments and study the particle dynamics in terms of the
ensemble and time averaged MSDs. We further analyse the non-Gaussianity of 
particle trajectories, the van Hove position correlation function, as well 
as the first passage statistics of the tracer.

The paper is organised as follows. In the next section we set up the model, discuss
the simulation procedure and the data analysis. In sections \ref{sec-results} and
\ref{sec-results1} we describe the main results for the homogeneous and
heterogeneous cases, respectively, and compare them to theoretical models. Further
analyses of both cases in terms of the non-Gaussianity parameter and
the van Hove correlation function are presented in section \ref{sec-results2}
while the first passage statistics are discussed in section \ref{sec-results3}. 
In section \ref{sec-discussion} we draw our conclusions and discuss some applications 
of our results.

\section{Model, simulations scheme, and data analysis}
\label{sec-model}

\begin{figure}
\begin{center}
\includegraphics[width=12cm]{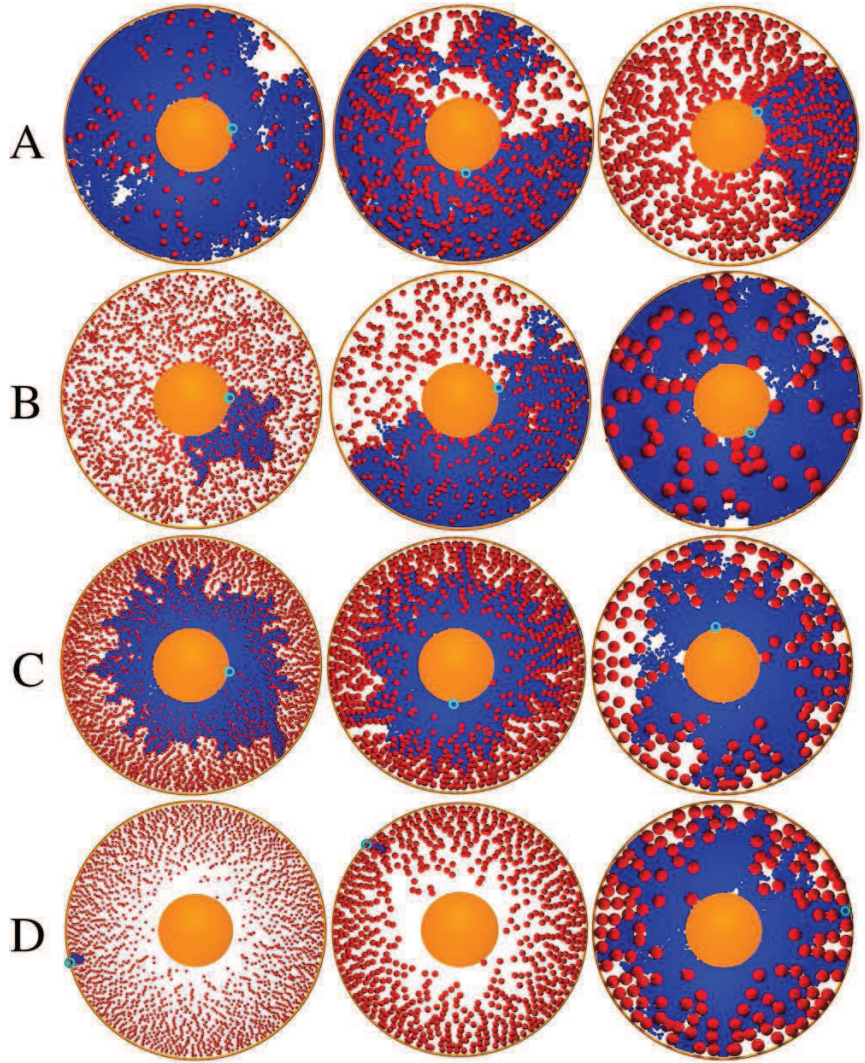}
\end{center}
\caption{Diffusion profiles for homogeneously (panels A and B) and heterogeneously
(panels C and D) distributed monodisperse crowders. In panel A the fraction of
crowders increases from left to right: $\phi=0.05$, $0.2$, $0.3$, and the size of
the crowders is fixed to $R_c=2$. In panel B the size of the crowders grows from
left to right: $R_c=1$, $2$, $5$, while the crowding fraction is fixed at $\phi=
0.2$. For heterogeneous crowders distributions in panels C and D, the crowding 
fraction $\phi(r)$ exhibits the linear growth (\ref{eq-rho-sub}) in radial
direction. The cell nucleus is represented by the
excluded disk at the centre, shown in orange. The diffusion is in-out (i.e., from
the nucleus to the membrane) in panel C and out-in (from the membrane to the
nucleus) in panel D. The initial tracer position is shown by the small blue
circle. The crowder radius in panels C and D grows from left to right:
$R_c=1$, $2$, $5$. The regions of
the domain visited by the tracer up to the diffusion time $T=10^4$ are depicted
in blue. Major asymmetries arise at higher crowding fraction and in heterogeneous
systems.}
\label{fig-scheme}
\end{figure}

We consider a model cell in the form of a planar circular annulus between the 
nucleus, represented by the excluded region within the radius $a$, and the plasma 
membrane located at radius $R$, as shown in figure \ref{fig-scheme}. The space between 
the membrane and the nucleus is filled with \emph{static\/} monodisperse crowders 
of radius $R_c$.  We consider two cases: homogeneous distribution of crowders with 
a prescribed crowding (area) fraction $\phi$, and heterogeneous distribution of 
crowders with a linear radial gradient
\begin{equation}
\phi(r)=\phi(a)+\frac{\phi(R)-\phi(a)}{R-a}(r-a)
\label{eq-rho-sub}
\end{equation} 
for $a<r<R$. For most cases we set $\phi(a)=0.01$ and $\phi(R)\approx0.3$. As we show below,
equation (\ref{eq-rho-sub}) leads to a transiently subdiffusive tracer motion
from the nucleus to the cell membrane, emerging due to an increasing density
of crowders near the cell periphery, as evidenced by figure \ref{fig-scheme}, panels
C and D.

In the simulations the crowders are placed at random positions without overlap. The
highest crowding fraction we simulated was around 30\%. When computing the mean time
averaged MSD in equation (\ref{eq-tamsd-average-2}) below, we typically average over
$M =10^2$ random configurations of crowders. This disorder average is taken in
addition to the average over individual trajectories in a given, quenched crowder
configuration. The tracer particle has a fixed unit radius. We set the radius of
the circular membrane to $R=100$ and the radius of the inner nucleus to $a=30$, so
that the ratio $a/R$ is similar to that of a typical eukaryotic cell \cite{lang11}.
The radii in the simulations and in the plots are measured in terms of the
length scale $\sigma$ of the potential (\ref{eq-lj}).
% $\sigma = 6$~nm.

The Weeks-Chandler-Andersen repulsive potential given by the 6-12 Lennard-Jones
potential $E_{\mathrm{LJ}}(r)$ with the standard cutoff distance $r_{\mathrm{cut}}$
is used to parameterise the repulsion between the tracer and crowders,
\begin{equation}
E_{\mathrm{LJ}}(r)=4\epsilon\left[\left(\frac{\sigma}{r}\right)^{12}-\left(\frac{
\sigma}{r}\right)^6+\frac{1}{4}\right] \label{eq-lj}
\end{equation} 
for $r<r_{\mathrm{cut}}=2^{1/6}\sigma$, and $E_{\mathrm{LJ}}(r)=0$ otherwise
\cite{wca71}. We simulate the dynamics of
the centre position $\textbf{r}(t)$ of the tracer via the Langevin equation
\begin{eqnarray}
\nonumber
m\frac{d^2\textbf{r}(t)}{dt^2}&=&-\sum_J\boldsymbol\nabla\Big[E_{\mathrm{LJ}}
(|\textbf{r}-\textbf{R}_{J}|-(R_c+\sigma))+E_{\mathrm{LJ}}(|\textbf{r}|
-(a+\sigma))\\ &&+E_{\mathrm{LJ}}((R-\sigma)-|\textbf{r}|)\Big]-\xi\textbf{v}(t)+
\textbf{F}(t),
\label{eq-langevin}
\end{eqnarray} 
where $m$ is the mass of the tracer particle, $\xi$ is the friction coefficient 
experienced by the tracer particle, $\textbf{v}(t)$ is its velocity, and 
$\textbf{{R}}_{J}$ is the static position of the $J$th crowder. Finally 
$\textbf{{F}}(t)$ represents a Gaussian $\delta$ correlated noise with 
zero mean and covariance
\begin{equation}
\left<\textbf{{F}}_j(t)\cdot\textbf{{F}}_k(t')\right>=2\delta_{j,k}\xi k_B
\mathscr{T}\delta(t-t').
\end{equation}
The inertial term in equation (\ref{eq-langevin}) gives rise to a ballistic regime
in the particle dynamics, as shown below (see reference \cite{ghos15} for
more details on this regime).  In the simulations below we set $\epsilon=k_B
\mathscr{T}=1$, $m=1$, and $\xi=1$ correspondiing to moderate damping
\cite{duko,forr,chen}. At these scales, the diffusivity of a tracer in an uncrowded
envinronment is $D_0=k_B\mathscr{T}/\xi=1$. The unit time step of simulations
corresponds to the physical time $\tau=\sigma\sqrt{m/(k_B\mathscr{T})}$. 
We employ the Verlet velocity algorithm with the time step $\delta t=0.01$ to 
integrate the stochastic equation (\ref{eq-langevin}). 

To characterise the diffusion behaviour we evaluate the time averaged
MSD
\begin{equation}
\overline{\delta^2(\Delta)}=\frac{1}{T-\Delta}\int_0^{T-\Delta}
\Big[\textbf{r}(t+\Delta)-\textbf{r}(t)\Big]^2dt
\label{eq-tamsd-average}
\end{equation}  
for individual particle trajectories $\textbf{r}(t)$. Here $T$ is the length of
the trajectory (observation time) and $\Delta$ is the lag time defining the width
of the window slid along the trajectory. This definition of the
time average is standard in single particle tracking experiments \cite{golding,
pt,metz14,aust06}. For $N$ individual trajectories the ensemble average 
is approximated as
\begin{equation}
\left<\overline{\delta^2(\Delta)}\right>=\frac{1}{N}\sum_{i=1}^N\overline{
\delta^2_i(\Delta)} .
\label{eq-tamsd-average-2}
\end{equation}
In the scenario of quenched heterogeneous environments considered herein we 
also calculate the disorder
average of the time averaged MSD over $M$ different realisations of the
crowding environment (compare \cite{sinai,yousof})
\begin{equation}
\label{distamsd}
\widetilde{\left<\overline{\delta^2(\Delta)}\right>}=\frac{1}{M}\sum_{j=1}^M
\left<\overline{\delta^2(\Delta)}\right>_j.
\end{equation}
The ensemble averaged MSD is also computed as double average over $N$
tracer trajectories for each crowders distribution and $M$ crowders
distributions:
\begin{equation}
\langle \mathbf{r}^2(t)\rangle = \widetilde{\left< [\mathbf{r}(t) - \mathbf{r}(0)]^2 \right>} = 
\frac{1}{MN}\sum_{j=1}^M \sum_{i=1}^N  [\mathbf{r}_i(t) - \mathbf{r}_i(0)]^2_j .
\end{equation}

\section{Homogeneous crowding case: ensemble and time averaged mean squared
displacements}
\label{sec-results}

\begin{figure}
\begin{center}
\includegraphics[width=14cm]{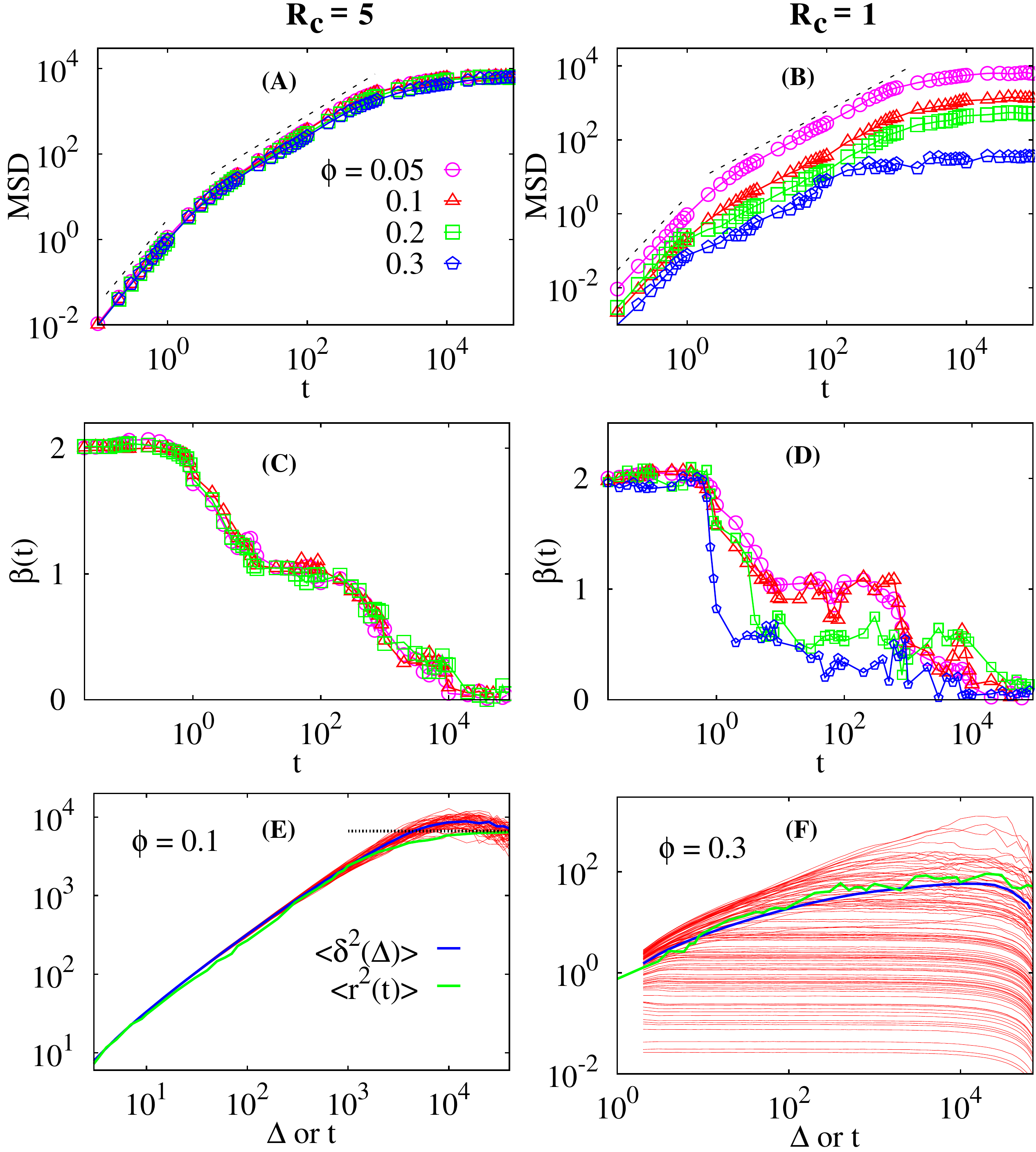}
\end{center}
\caption{MSD, local scaling exponent $\beta(t)$, and time averaged MSD
$\left<\overline{\delta^2}\right>$ for tracer diffusion in an
environment of homogeneously distributed crowders. The two columns correspond
to different crowder radii: $R_c=5$ (left) and $R_c=1$ (right). The data for
different crowding fractions $\phi$ are shown by different symbols in the
panels A, B, C, and D. In panels A and B the two dashed lines indicate the
ballistic asymptote at short times and the linear Brownian growth at intermediate
times. In panels E and F  100 individual time averaged MSD curves for a
particular distribution of crowders are plotted.
In panel E the long time plateau (\ref{eq-msd-plateau}) of the MSD is
shown by the dotted line. Parameters: the radius of the domain is $R=100$ and the radius of
the excluded nucleus region is $a=30$. The crowding fraction $\phi$ is indicated
in the panels. Where applicable, the number of different crowders configurations
used in the averaging for the MSD and the scaling exponent is $M=100$ (disorder
averaging), the number of tracer trajectories in each crowders configuration is $N = 100$ so that 
the total number of tracers used for averaging is $M\times N=10^4$.}
\label{fig-msd-homo}
\end{figure}

We start with the analysis of the tracer diffusion among static homogeneously
distributed crowders as shown in figure \ref{fig-scheme} in panels A and B.
Figure \ref{fig-msd-homo} shows the ensemble averaged MSD $\langle\mathbf{r}^2
(t)\rangle$ and the corresponding local MSD scaling exponent
\begin{equation}
\beta=\frac{d\log\langle\mathbf{r}^2(t)\rangle}{d\log t}.
\end{equation}
Both the crowding fraction $\phi$  and the crowder radius $R_c$ are varied.

For relatively large crowders an initial ballistic growth of the MSD (corresponding 
to underdamped particle motion, see also \cite{ghos15}) crosses over to a quite 
prolonged Brownian regime with scaling 
exponent $\beta\approx1$, as seen in panels A and C of figure \ref{fig-msd-homo}.
For the smaller crowders this effect is less pronounced, compare panels B and D.
At later times the tracer motion starts to be affected by the confinement exerted
by the outer reflecting membrane of the domain at $r=R$, and the MSD begins to
saturate to a plateau. Concurrently the scaling exponent $\beta(t)$ tends to zero.

The effect of the crowding fraction $\phi$ on the MSD behaviour is illustrated
in panels A and B of figure \ref{fig-msd-homo} for large and small crowders,
respectively. For a tracer of unit size the larger crowders do not appear
to create substantial obstruction for the simulated crowding fractions. Even
at relatively large $\phi$ values the MSD only marginally diminishes with
increasing crowding fraction $\phi$, see panel A. In other words, small tracers
always manage to manoeuvre around large void spaces remaining between large crowders.

In contrast, relatively small crowders at identical crowding fractions
$\phi$ yield severe reductions of the average diffusive tracer motion, as evidenced
by panel B of figure \ref{fig-msd-homo}. This observation is consistent with the
dramatic differences of the spatial exploration patterns exhibited by the tracer
in panels B of figure \ref{fig-scheme}. The magnitude of the plateau of the MSD
attained for larger $\phi$ values is much smaller than expected for an annulus
without crowders, compare with equation (\ref{eq-msd-plateau}) below. This
fact is due to the intricate labyrinthine environment formed by the unit sized
crowders for diffusion of a tracer of the same size. For small crowders the
value of the scaling exponent $\beta$ is reduced significantly and at much earlier
times for larger crowding fractions $\phi$, see panels C and D of figure
\ref{fig-msd-homo}. These features obviously strongly depend on the specific
quenched environment, in which the tracer motion occurs, resulting in a high degree
of irreproducibility of the tracer diffusion for different realisations of the
disorder. This is also quantified in panel F of figure \ref{fig-msd-homo} (see
below) and leads to substantially larger uncertainties in the local scaling
exponent $\beta(t)$ computed from the available MSD curves.   
The strong effect of small crowders on the MSD and the minor effect 
of the large crowders is our first main result.

We now turn to the analysis of the time averaged MSD obtained from averaging over
$10^4$ tracer trajectories for a single realisation of the crowder
configuration. As shown in panel E of figure
\ref{fig-msd-homo} for large crowders with $R_c=5$ and small crowding fractions
$\phi$ the spread of amplitudes of the time averaged MSD curves is quite small.
At later times, when the tracer motion starts to be influenced by the outer boundary
the average $\left<{\overline{\delta^2(\Delta)}}\right>$
approaches a plateau which has about twice the amplitude of the plateau value
of the ensemble averaged MSD, see panel E in figure \ref{fig-msd-homo}. Note that
because of the relatively small domain size used in the simulations, the moderate
trajectory lengths, and the presence of randomly distributed crowders this plateau
is not as distinct as, for instance, for the deterministic, confined HDP, compare
figure 4 in reference \cite{cher14c} and figure 8 in reference \cite{cher15a}.

We recall that for the uncrowded case the long time (plateau) values of
the ensemble and time averaged MSDs are related to the inner and
outer radii in two dimensions as 
\begin{equation}
\left<x^2\right>_{\mathrm{st}}=\frac{1}{2}\left<\overline{\delta^2}
\right>_{\mathrm{st}}\approx\frac{R^2+a^2}{2}.
\label{eq-msd-plateau}
\end{equation}

The occurrence of the factor $1/2$ for the time averaged MSD is inherent to the
very definition (\ref{eq-tamsd-average}) \cite{metz14,greb11a,pre12}. The asymptote
(\ref{eq-msd-plateau}) is shown in panel E of figure \ref{fig-msd-homo} by the
dotted line. The attainment of a plateau value of both ensemble and time averaged
MSDs on a bounded domain is a typical feature of both ergodic processes such
as Brownian motion and fractional Brownian motion as well as weakly non-ergodic
processes, and is thus inherently different from the deviations from a plateau
value in confining potentials \cite{metz14,pre12}.

In contrast to these observations, for small crowders at high crowding fractions
($\phi=0.3$) the amplitude spread of the time averaged MSD curves for a particular
distribution of crowders is quite pronounced, as shown in panel F of figure
\ref{fig-msd-homo}. The magnitude of the time averaged MSD is much smaller than
that for larger crowders, compare the magnitude of the time averaged MSD in panels
E and F of figure \ref{fig-msd-homo}. In fact, time averaged MSD curves with 
very small magnitudes (below $1$) resulted from almost immobile finite size
tracers that were blocked by surrounding crowders.
Moreover, the tracer only rarely reaches the
outer cell membrane but mostly saturates at much lower values due to confinement
by the crowders in a subdomain of our model cell. The MSD itself features a much
more pronounced
amplitude scatter for the same number $N$ of traces used in the averaging.
Because of the tracer localisation and the wide amplitude spread of the time
averaged MSDs the mean time averaged MSD $\left<\overline{\delta^2}\right>$ has
poor statistics and its relation to the ensemble averaged MSD prescribed by
equation (\ref{eq-msd-plateau}) is difficult to check. This behaviour of the
time averaged MSD in homogeneous crowding environments is our second main result.

\section{Heterogeneous crowding case: ensemble and time averaged mean squared
displacements}
\label{sec-results1}

How does a heterogeneous distribution of crowders affect the above results? 
An immediate effect of the heterogeneous distribution consists in
very different properties for the in-out (from the nucleus
surface to the membrane) and out-in (from the membrane to the nucleus) scenarios
of the tracer diffusion. The average local density of crowders in the simulation
domain (see panels C and D of figure \ref{fig-scheme}) is generated according 
to equation (\ref{eq-rho-sub}) which naturally leads to a higher local
diffusivity close to the nucleus that corresponds to the experimentally relevant
diffusion of small proteins inside surface adhered eukaryotic cells \cite{lang11,cher14a}.

\begin{figure}
\begin{center}
\includegraphics[width=14cm]{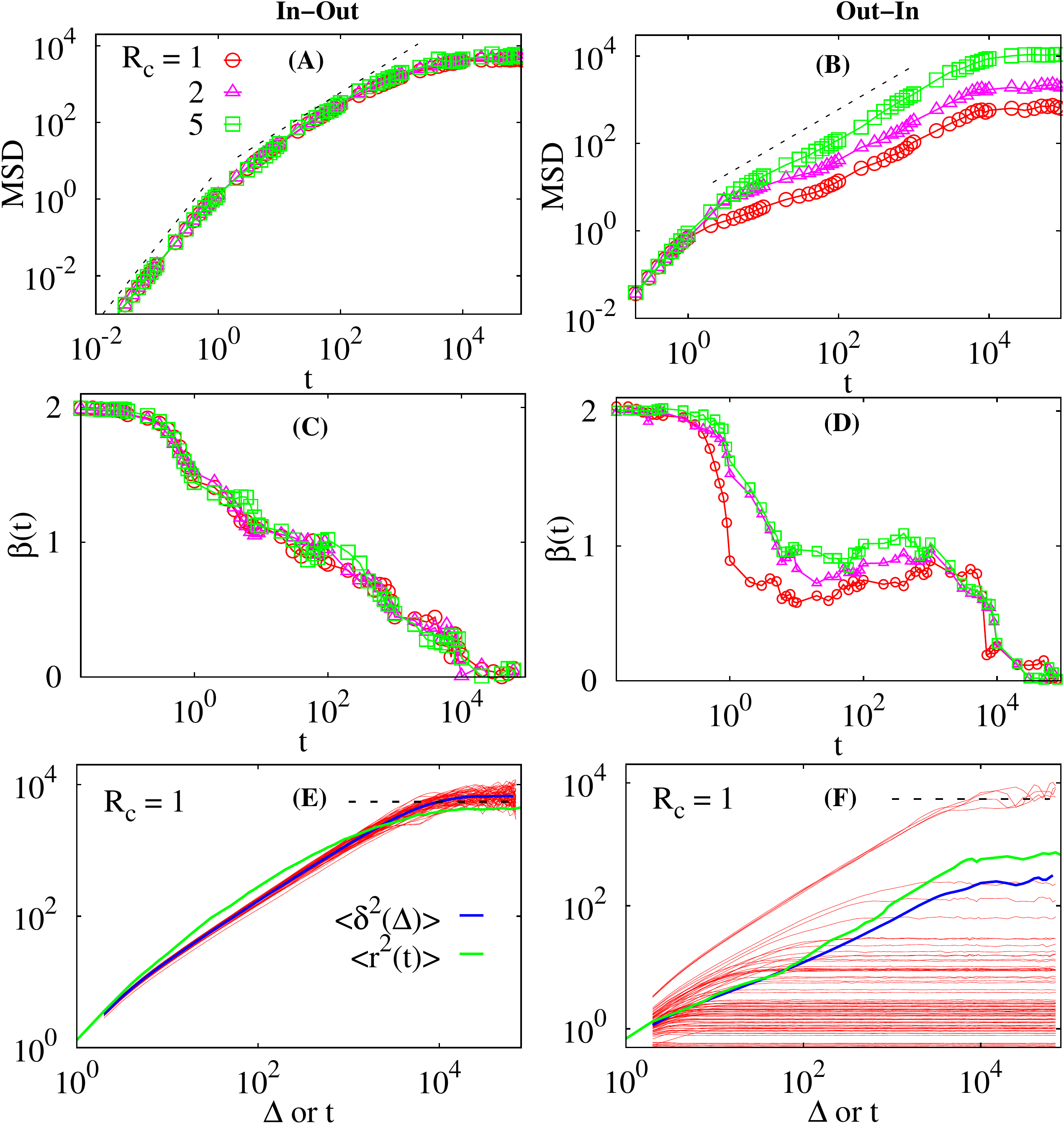}
\end{center}
\caption{MSD, local scaling exponent $\beta(t)$, and time averaged MSD
for tracer particles diffusing in heterogeneous crowder distributions of the
form (\ref{eq-rho-sub}). The two columns correspond to the in-out and out-in
tracer diffusion, that is, respectively, for the release of the tracer particle
at the inner and outer radii of the model cell. The data for different crowder
radii $R_c$ are shown by different symbols in panels A-D. In panels A and B the
dashed lines indicate the ballistic asymptote at short times and the linear
Brownian regime at intermediate times. In panels E and F the long time plateau
of the MSD for the uncrowded case given by the asymptote (\ref{eq-msd-plateau}) 
is shown by the dashed line. 
Parameters: $R=100$, $a=30$, $T=10^5$, $M = 100$, and $N=100$.}
\label{fig-msd-hetero}
\end{figure}

For the in-out diffusion from the nucleus to the cell periphery the
tracers become trapped in progressively denser arrangements of crowders.  
At the same crowding fraction $\phi$, these trapping situations arise 
earlier in time for smaller crowders as compared to larger crowders, compare 
the panels C in figure \ref{fig-scheme}. Similar to the homogeneous case the MSD
starts ballistically, then reveals a linear Brownian regime, and finally saturates
due to the confinement in the annulus. Concurrently, the local scaling
exponent $\beta(t)$ decreases continuously and finally vanishes when the
MSD approaches a plateau, see panels A and C of figure \ref{fig-msd-hetero}.
The spread of the time averaged MSD curves is relatively
small and their long time plateau is again about twice of that of the ensemble
averaged MSD for a particular crowder configuration, as expected from relation
(\ref{eq-msd-plateau}) and seen in panels
E of figures \ref{fig-msd-homo} and \ref{fig-msd-hetero}. At intermediate times
a stronger disparity between the ensemble and time averaged MSD is observed when
compared to the homogeneous case, as shown in panel E of figure
\ref{fig-msd-hetero}. The scatter of the time averaged MSD curves
decreases when longer trajectories (larger $T$ values) are analysed (results not
shown). This behaviour is expected and is realised for several ergodic and nearly
ergodic processes \cite{metz14,greb11a}. The magnitude of the amplitude scatter of the
time averaged MSDs in panels E of figures \ref{fig-msd-homo} and
\ref{fig-msd-hetero} is similar to that of Brownian motion \cite{metz14,greb11b,andr}.

In the opposite case of out-in diffusion (panel D of figure \ref{fig-scheme}) we
observe that for relatively high crowding fractions of small crowders
a finite size tracer often cannot even leave the vicinity of the boundary, giving
rise to prolonged trapping events in this confined area. This leads to a large
proportion of low amplitude, nearly constant time averaged MSD curves, as
seen in panel F of figure \ref{fig-msd-hetero}. The magnitude of the mean time
averaged $\left<\overline{\delta^2(\Delta)}\right>$ is in many cases
dominated by several successful fast translocation events of tracers from the cell
membrane to the nucleus. In the long time limit equation (\ref{eq-msd-plateau})
is thus not valid in this situation. The tracer localisation and the dominance of
one or few extreme tracer trajectories in the mean value $\left<
\overline{\delta^2(\Delta)}\right>$ is also a rather common feature of stochastic
processes in the presence of well pronounced traps as well as in ageing stochastic
processes \cite{metz14,johannes}. The distinctly different behaviour between 
out-in and in-out diffusion in our heterogeneous crowder system is the third main
results of this study.

Since the crowding fraction in equation (\ref{eq-rho-sub}) grows from the nucleus towards 
the cell periphery, the in-out diffusion in such heterogeneous crowders distribution 
is expected to be subdiffusive \cite{cher14a}. In this scenario the tracers are progressively
trapped closer to the cell periphery. Here, however, we observe the formation of a
radial percolation in the circular domain: the tracers are not allowed to penetrate beyond
some critical radius that features a particular critical density of crowders, an
effect that is crucially related to the finite size of both the crowders and the
tracer particle. 
Similar effects of local confinement are naturally observed for single
particle trajectories of tracer diffusion in random percolation systems
\cite{yousof}. The initial period of the tracer diffusion that occurs in the
region with low crowder density is naturally reproducible and leads to a
small spread of the time averaged MSD curves, as shown in panel E of figure
\ref{fig-msd-hetero}.  

In contrast, for the out-in diffusion the tracer starts in the region of the highest
concentration of crowders and diffuses into regions containing less and less
crowders. The associated initial tracer localisation events give rise to stalling
time averaged MSD curves with a very large spread, see
panel F of figure \ref{fig-msd-hetero}. Early trapping in this scenario leads to
the emergence of a longer plateau region in the corresponding time averaged MSD
curve. The mean time averaged MSD computed over an ensemble of $N$ trajectories
is rather imprecise because of the dominance of few extreme events.

\begin{figure}
\includegraphics[width=16cm]{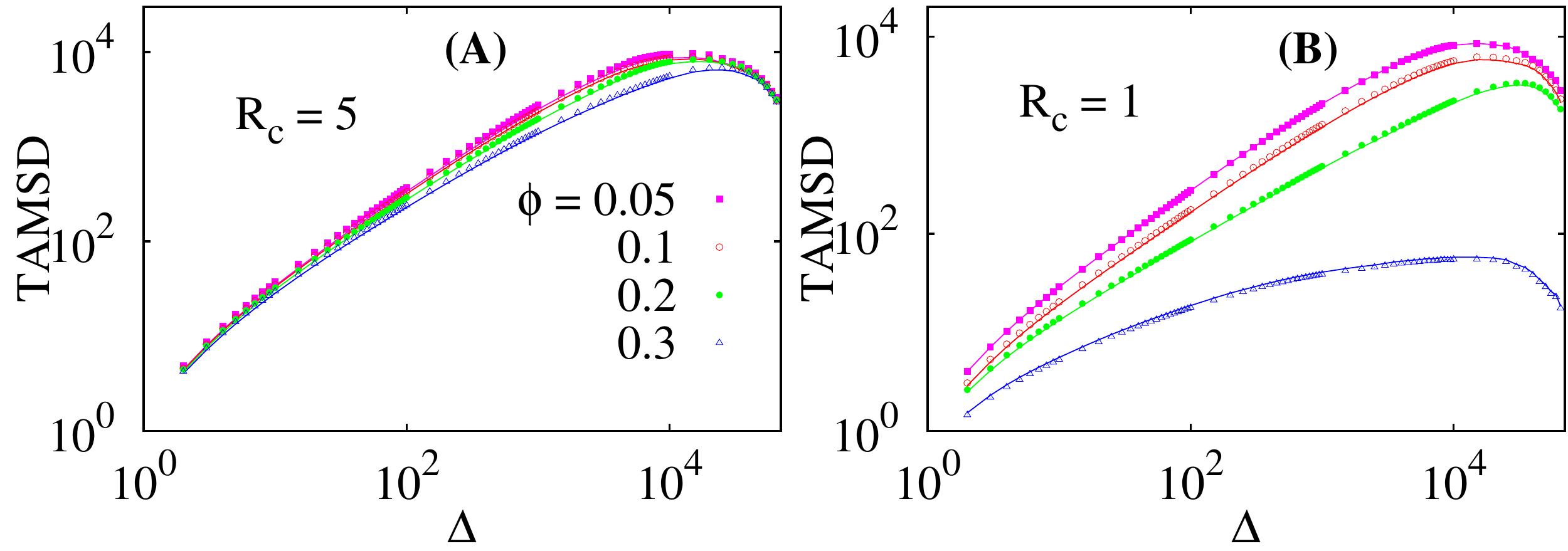}
\caption{Time averaged MSD curves for trajectory based averaging only 
(solid lines; $M=1$, $N=10^4$) and for double averaging over tracer 
trajectories and realisations of the crowder distributions 
(symbols; $M=10^2$, $N=10^2$), computed for homogeneously
distributed crowders. These curves can be compared to the results 
in panels E and F of Fig. \ref{fig-msd-homo}. 
The crowder radius $R_c$ and the crowding fraction
$\phi$ are indicated in the plots. Note that individual time 
averaged MSD curves are not shown, only the averages
$\left<\overline{\delta^2}\right>$ and $\widetilde{\left<\overline{\delta^2}
\right>}$ are presented.}
\label{fig11}
\end{figure}

Figure \ref{fig11} compares the averaging over $N$ different tracer trajectories
starting at random points at the nucleus boundary for a single
crowder distribution and the double averaging that includes the average over $M$
random realisations of crowders distribution. The time averaged MSDs for $N=100$ 
and $M=100$ are shown by symbols, while those for the case $M=1$ and $N=10^4$ are shown as solid lines. 
The two cases of homogeneously distributed crowders for large
(panel A) and small (panel B) crowders are presented. We observe that
the differences between the two averaging approaches are rather small: the single
average over $10^4$ trajectories and the double average over $10^2$ trajectories and $10^2$
crowder distributions yield very similar results. 

\begin{figure}
\begin{center}
\includegraphics[width=14cm]{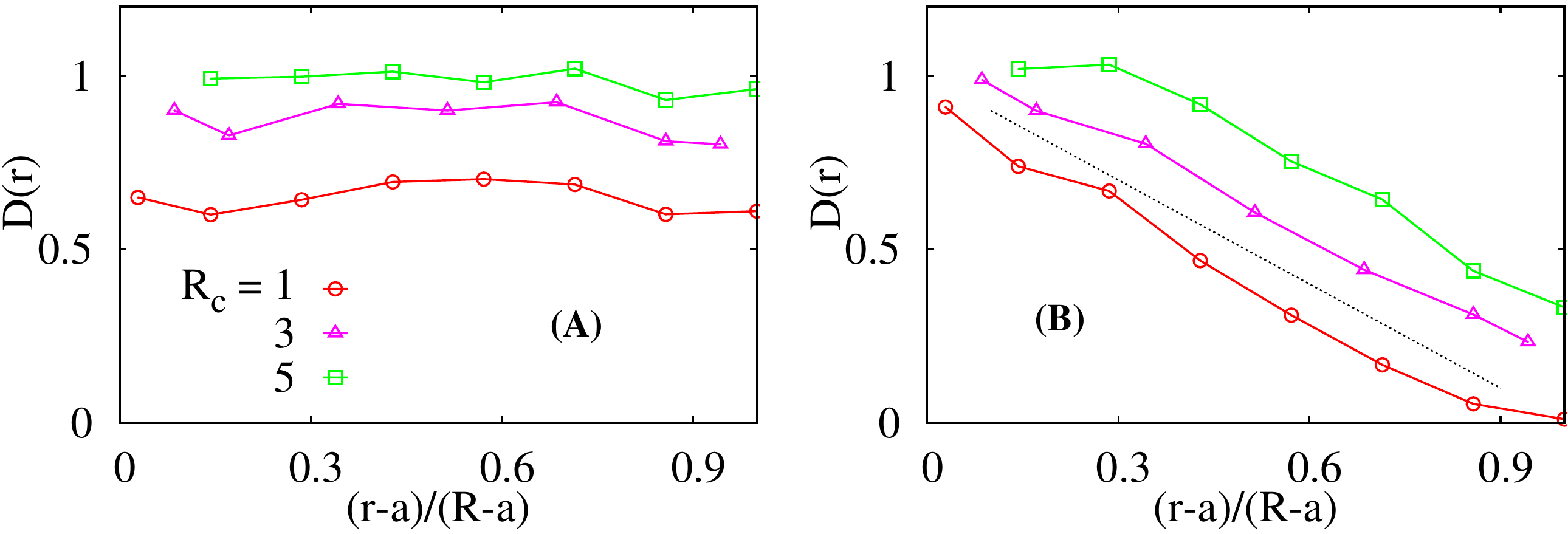}
\end{center}
\caption{Effective radial diffusivity of tracers for (A) homogeneous ($\phi=0.1$)
and (B) heterogeneous ($\phi(r)$) crowder distributions. Other parameters are the
same as in figures \ref{fig-msd-homo} and \ref{fig-msd-hetero}, respectively, and
the crowder radii $R_c$ are as indicated. The dashed line in panel (B) represents
the asymptote (\ref{eq-diff-linear}) without the last constant term.}
\label{fig-dr}
\end{figure}

We checked that for homogeneously distributed crowders the radial tracer diffusivity
is approximately constant (panel A of figure \ref{fig-dr}), as it should be.  
In turn, for a heterogeneous distribution of crowders given by
equation (\ref{eq-rho-sub}) the radial tracer diffusivity is a decreasing and
nearly linear function of the distance from the nucleus (panel B of
figure \ref{fig-dr}),
\begin{equation}
D(r)\approx D_0 \left(1 - 0.78 \frac{r-a}{R-a}\right) + \mathrm{const}.
\label{eq-diff-linear}
\end{equation}
The slope $0.78$ is nearly independent on the crowder radius $R_c$. This
linear dependency of the diffusivity seems to be a consequence of the linear 
increase of the crowders density as given by equation
(\ref{eq-rho-sub}). This is our fourth main result.

\section{Non-Gaussianity parameter and van Hove correlation function}
\label{sec-results2}

\begin{figure}
\begin{center}
\includegraphics[width=14cm]{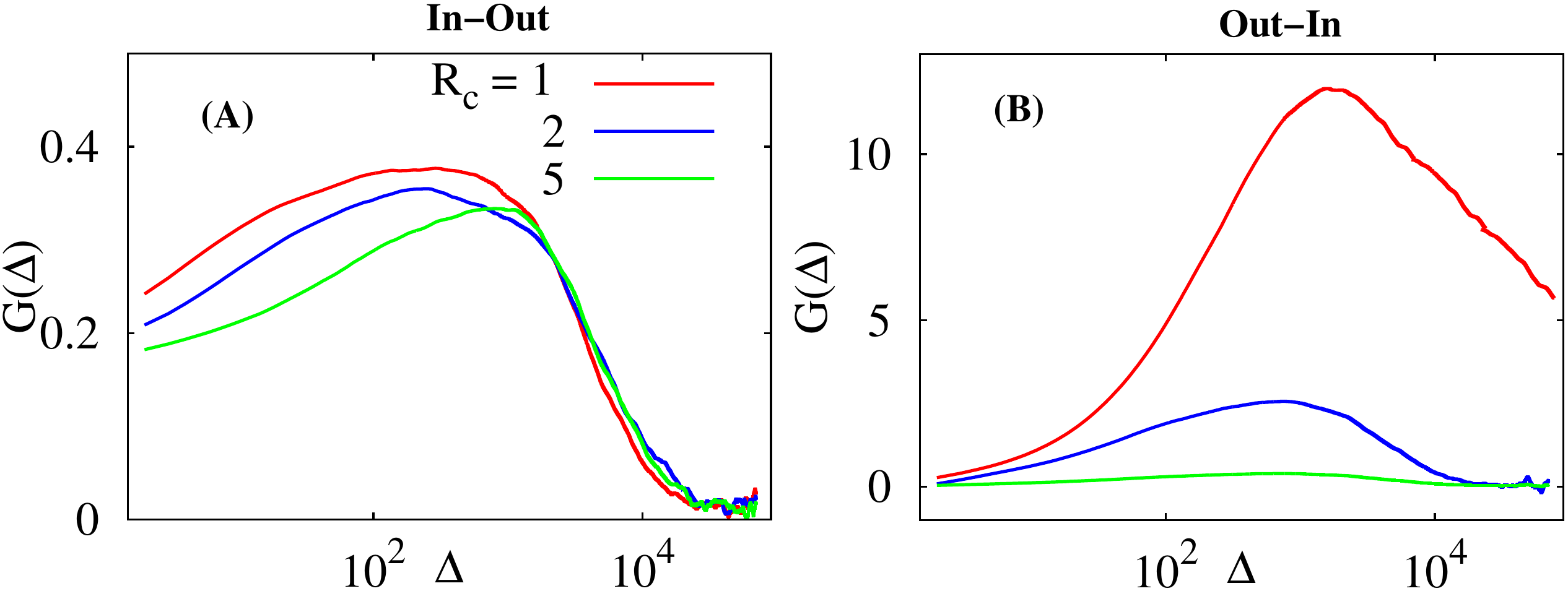}
\end{center}
\caption{Non-Gaussianity parameter $G(\Delta)$ (\ref{eq-nong}) for in-out and out-in
tracer diffusion. Parameters are the same as in figure \ref{fig-msd-hetero}
averaged over $N=10^4$ tracer trajectories for a single configuration
of crowders ($M = 1$).}
\label{fig-nong}
\end{figure}

Following reference \cite{ghos15} we now proceed to evaluate
the experimentally relevant non-Gaussianity parameter for the tracer diffusion in
our crowded environment. Similar to the ergodicity breaking parameter EB
\cite{metz14,he} it contains the fourth order moment of the tracer time averaged
MSD averaged over $N$ realisations. Namely, in two dimensions we have
\cite{fran13,voth06}
\begin{equation} 
G(\Delta)=\frac{\left<\overline{\delta^4(\Delta)}\right>}{2\left<\overline{
\delta^2(\Delta)}\right>^2}-1.
\label{eq-nong}
\end{equation}

We find that for in-out diffusion the non-Gaussianity parameter assumes moderate
values for shorter lag times $\Delta$ while it becomes close to zero for longer
lag times, as seen in panel A of figure \ref{fig-nong}. This is a typical long time behaviour
of ergodic tracer diffusion, compare figure 3a in reference \cite{voth06}.
Indeed, as we show in panel E of figure \ref{fig-msd-hetero} in the long time limit
the ensemble and time averaged MSDs differ simply by the above mentioned factor of
2 for heterogeneous crowder distributions. In contrast for out-in diffusion the
non-Gaussianity parameter
attains substantially larger values. This feature is likely due to the highly
non-reproducible trajectories of the tracer motion and prolonged localisation
events near the cell periphery in the region of high crowder concentration.

The trapping as well as the non-Gaussianity of the tracer
diffusion can also be characterised by the van Hove correlation function $G_s(\Delta x,\Delta t)$
describing the probability that a particle moves a distance $\Delta x$ during
time $\Delta t$ \cite{skau14,voth04},
\begin{equation}
G_s(\Delta x,\Delta t)=\frac{1}{N}\sum_{i=1}^{N}\delta\Big(x_i(\Delta t)-x_i(0)-
\Delta x\Big),
\label{eq-hove-function}
\end{equation} 
where $N$ is the number of tracers used for averaging. For a system of hard spheres
the van Hove correlation function simply corresponds to the diffusion propagator
\cite{schm10} governed by the diffusion equation.

\begin{figure}
\begin{center}
\includegraphics[width=14cm]{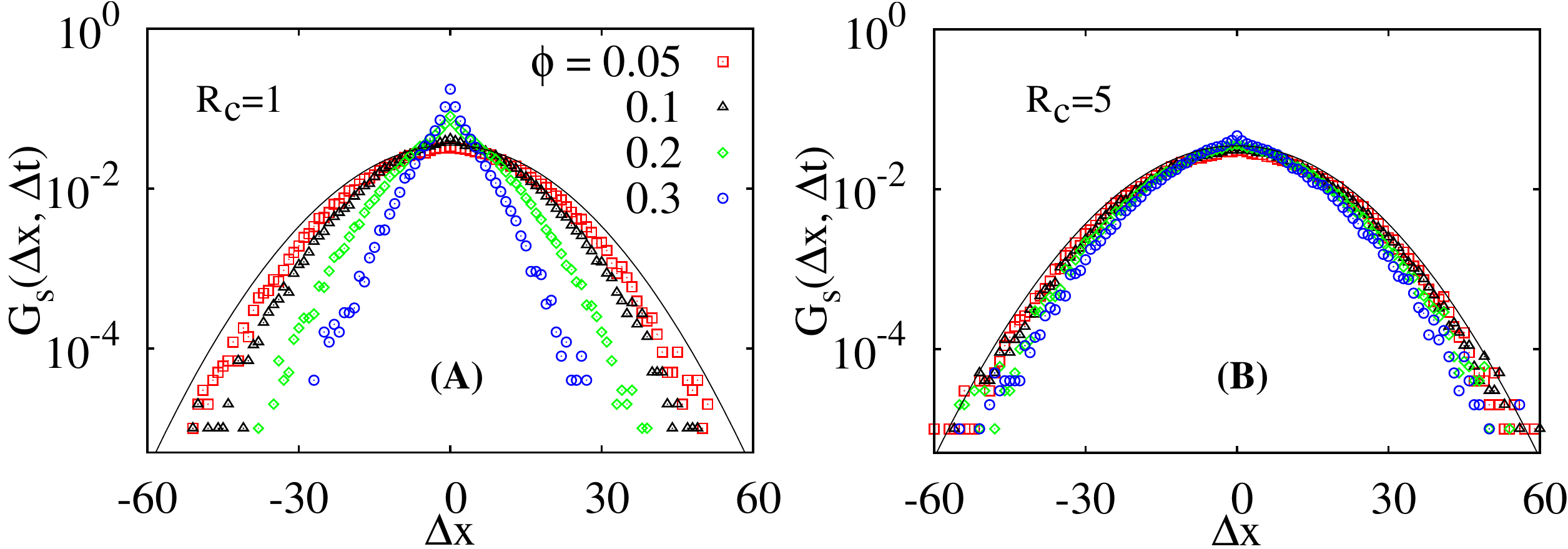}
\end{center}
\caption{Van Hove correlation function for homogeneous crowder distribution
and several crowding fractions $\phi$. Other parameters are the same as in figure
\ref{fig-msd-homo}. The solid line shows the approximately Gaussian function obtained
from our simulations for a tracer restricted to an annulus without crowders.
Panels A and B are for crowder radii $R_c=1$ and $R_c = 5$, respectively. }
\label{fig-hove}
\end{figure}

Figure \ref{fig-hove} shows the behaviour of the van Hove correlation function for
tracer diffusion among homogeneously distributed crowders. We observe
that for relatively large crowders the probability of trapping events of the
tracer particle is quite low and the distribution of tracer displacements remains
close to Gaussian for all crowding fractions studied, as it should (see panel B
in figure \ref{fig-hove}). In contrast, for small crowders the non-Gaussianity of
the van Hove function becomes quite pronounced, in particular for larger crowding
fractions $\phi$, as demonstrated in panel A of figure \ref{fig-hove}. A faster decay of
the tracer displacements at high crowding fractions $\phi$ and for small crowders
is a consequence of tracer caging by crowders and anomalously slow diffusion. The
almost exponential distributions for high crowding fraction in panel A of figure \ref{fig-hove} 
compare well with the experimentally
measured step size distributions for polymer diffusion on nano-patterned surfaces
presented in figure 4 of reference \cite{schw15} as well as for liposome diffusion
in nematic solutions of actin filaments in figure 3C of reference \cite{gran12}. 
Heterogeneously structured environments for tracer diffusion cause a separation of 
particles into slow and fast populations (compare reference \cite{cher13b}),
reflected in a cusp of the particle distribution near the origin and longer than
Gaussian tails for large particle displacements.

\begin{figure}
\begin{center}
\includegraphics[width=14cm]{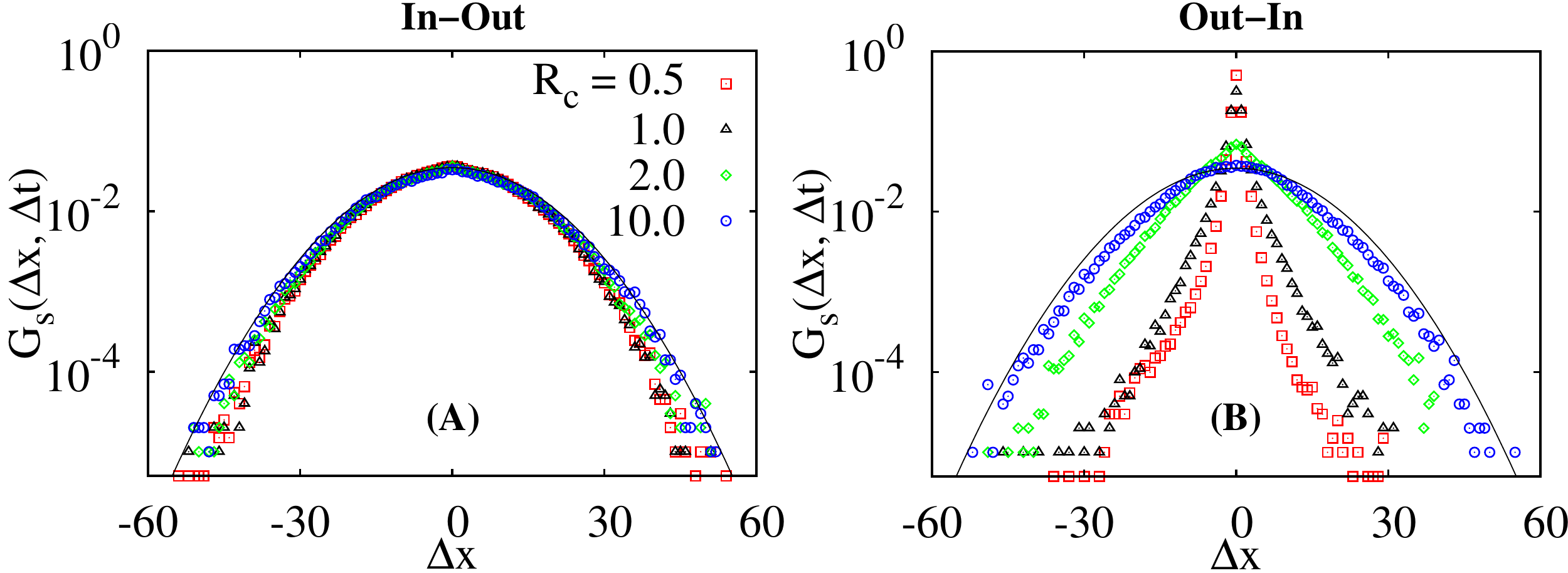}
\end{center}
\caption{Van Hove correlation function for (A) in-out and (B) out-in diffusion
plotted for several crowder radii $R_c$ in the scenario of heterogeneous crowders
distribution. The solid line shows the distribution of a tracer restricted to an annulus
without crowders: it almost superimposes with the results for the
largest crowders at $R_c=10$. Other parameters are the same as in figure
\ref{fig-msd-hetero}.}
\label{fig-hove-2}
\end{figure}

For heterogeneous distributions of crowders the behaviour of the van Hove function
is illustrated in figure \ref{fig-hove-2}. We observe that for in-out diffusion
the distribution remains approximately Gaussian for all crowder radii $R_c$ in our
simulations. This corresponds to the rather small non-Gaussianity parameter and
quite limited spread of the time averaged MSD curves in this situation. In contrast,
for out-in diffusion among small crowders the probability distribution of
tracer displacements
becomes progressively non-Gaussian. It features a pronounced cusp near $\Delta x=0$
describing a prevalence of small displacements characteristic for subdiffusive
processes in rather confined conditions. The detailed behaviour of the
non-Gaussianity parameter and the van Hove correlation function represents the
fifth main result of the present study.

\section{Survival probability} 
\label{sec-results3}

First passage time statistics are important to describe cellular processes, for
instance, to quantify the diffusion limit of reactions triggered by incoming, 
diffusing molecules. In the present cell model we focus on the first passage
behaviour of particles arriving to the membrane from the nucleus surface (in-out
case) or arriving to the nucleus surface from the cell boundary (out-in case).
To examine this behaviour we consider the survival probability $S(t)$ that a tracer
started either at the inner nucleus or at the outer boundary, and does not attain
a distance $(a+r)$ away from the cell centre up to time $t$. The survival
probability is directly related to the probability density of the first arrival
time of the tracer to that distance \cite{redn01,metz14c}.

\begin{figure}
\includegraphics[width=16cm]{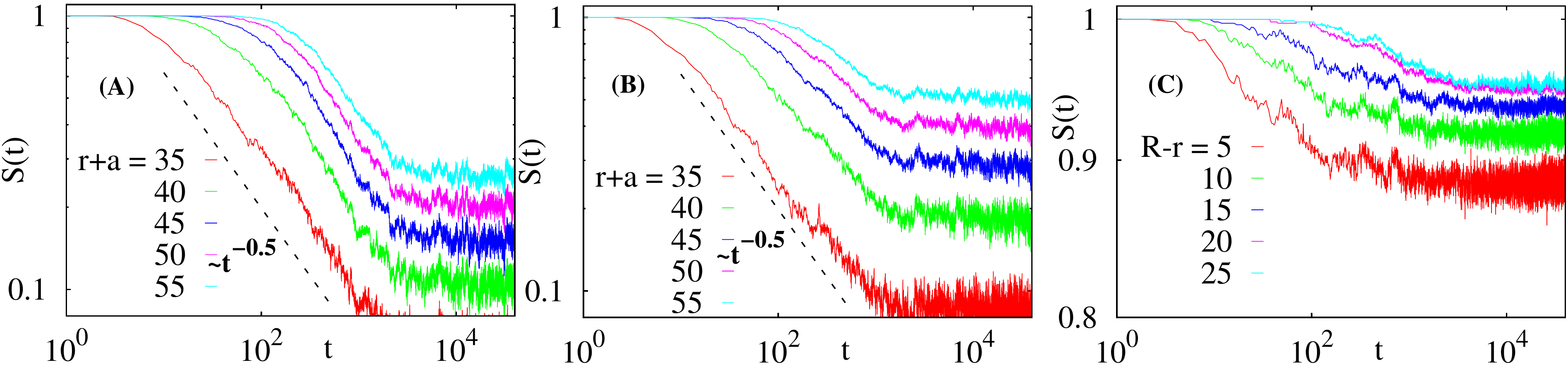}
\caption{Survival probability $S(t)$ for (A) homogeneous crowding (tracers are released near the 
nucleus), and for the heterogeneous crowding
described by equation (\ref{eq-rho-sub}) for the cases of (B) in-out and (C) out-in
diffusion. The asymptote $S(t)\sim t^{-1/2}$ from (\ref{eq-surv-scaling-in-out}) is shown in panels A
and B. Different colours indicate $S(t)$ for various distances $(r+a)$
from the cell centre within which the tracers are counted. Neither the inner
boundary at $r=a$ nor the outer boundary at varying radius $(a+r)$ are absorbing in
the simulations. The model parameters for homogeneous and heterogeneous crowder
configurations are the same as in figures \ref{fig-msd-homo} and
\ref{fig-msd-hetero}, respectively.}
\label{fig-surv}
\end{figure}

\begin{figure}
\includegraphics[width=16cm]{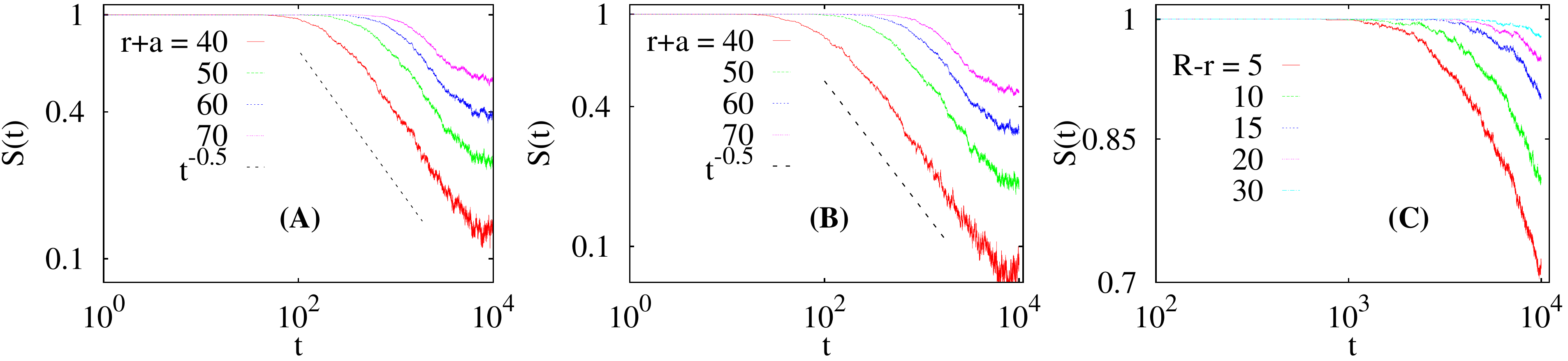}
\caption{The same as in figure \ref{fig-surv} but computed for the effective
radial diffusivity $D(r)$ from figure \ref{fig-dr}. The asymptote
$S(t)\sim t^{-1/2}$ is shown in panels A and B.}
\label{fig-surv-dr}
\end{figure}

Figure \ref{fig-surv} shows the survival probability $S(t)$ for
homogeneously and heterogeneously distributed crowders. For both homogeneous case and in-out diffusion we
observe at intermediate times
\begin{equation}
S(t)\simeq t^{-1/2}.
\label{eq-surv-scaling-in-out}
\end{equation}
In turn, for out-in diffusion $S(t)$ has only a very weak dependence on the diffusion
time $t$.
As the distance $(a+r)$ increases the survival probability starts to follow the scaling
behaviour at later times, see figure \ref{fig-surv}.
Similar scaling relations were obtained for subdiffusive
HDPs with a diffusivity of the form $D(r)=D_0 A/(A+r^2)$ for in-out diffusion of
tracer particles \cite{cher14a}. 
We also simulated the tracer diffusion in the annulus without crowders but with the 
effective radial diffusivity $D(r)$ presented in figure \ref{fig-dr}.
In these simulations, the same scaling law (\ref{eq-surv-scaling-in-out}) for the 
survival probability was obtained, compare figures \ref{fig-surv} and \ref{fig-surv-dr}.

Figure \ref{fig-surv} shows that the survival probability appears to saturate 
to finite values, instead of decaying to zero. The nonzero limiting value of the survival 
probability is related to the disorder averaging and should be equal to the fraction of 
crowders configurations, in which the circle of radius $a+r$ is not accessible
for a finite size tracer. In fact, for such configurations the survival probability is 
$1$, and it is then weighted by the fraction of these configurations.  The behaviour of 
the survival probability is the sixth main result of this study.

\section{Discussion and conclusions}
\label{sec-discussion}

The cytoplasm of living cells is a complex, superdense \cite{golding}
mixture of various molecules of highly variable sizes, shapes, and surface
properties, with often non-trivial spatial density distributions \cite{lang11}.
The passive diffusion of proteins such as transcription factors or
enzymes, as well as other complex signalling molecules in living cells represents
a vital ingredient in the cellular gene regulation and metabolism \cite{spako}.
For instance, transcription factors that are
taken up by the cell at its boundary need to diffuse to
the nucleus, in which they will control the information transfer of certain genes.
Conversely, viral components may be produced in the nucleus and are then
transported by thermal diffusion (while larger parts are actively by molecular
motors) to the cell membrane, where the assembly process of these viruses takes
place
\cite{zlot13}. The reverse process is, inter alia, relevant for the transport of
internalised viruses to the nucleus \cite{seis01,gode16}.  Recent experimental studies of the motion of
relatively small green fluorescent proteins (GFPs) in both the cytoplasm and the
nucleoplasm of living eukaryotic cells indeed demonstrate the existence of
subdiffusion up to the millisecond range \cite{gratton}, concurrent to the
existence of significant diffusivity gradients experienced by smaller proteins
\cite{lang11}.

In earlier works, the spatial heterogeneity was often modelled through a 
space-dependent diffusivity.  In particular, the implications of spatially 
heterogeneous diffusion processes
(HDPs) with a prescribed gradient of the diffusion coefficient $D(x)$ were studied
on the basis of the overdamped Langevin equation for different functional forms of
$D(x)$ \cite{cher13,cher14a,cher14c,cher15a}. These deterministic forms prescribe
a systematic variation of the local particle diffusivity, similar to diffusivity
maps in living cells \cite{lang11,elf11}, and are thus inherently different from 
spatially and temporally \emph{random} diffusivities \cite{slat14,lape14}. 
In particular, HDPs in a circular planar cell model with a radial 
diffusivity of the form $D(r)=D_0 A/(A+r^2)$ exhibit radial subdiffusion 
with an anomalous diffusion exponent $\beta=1/2$. In 
contrast to Brownian motion, the HDPs are weakly non-ergodic in the sense that 
time averages of physical observables such as the particle mean squared displacement 
do not converge to the corresponding ensemble averages even in the limit of long 
observation times \cite{metz14}.

In order to reveal the effect of molecular heterogeneity onto the 
intracellular transport, we performed extensive Langevin dynamics 
simulations of the motion of a passive tracer of a finite size in 
a two dimensional quenched disordered environment in which crowders are either 
homogeneously or heterogeneously distributed.
The simulation domain was an annulus, which was limited
by concentric inner and outer boundaries, representing the surface of the nucleus
and the plasma membrane of the cell. For varying crowding fraction and crowder size we
quantified the motion of the tracer in the cell in terms of the
diffusion profiles, the ensemble averaged MSD and the associated scaling exponent,
as well as the time averaged MSD and the amplitude scatter observed for individual
tracer trajectories.
Finally, we also determined the degree of the non-Gaussian behaviour
and the van Hove correlation function, as well as the survival statistics
of the tracer. 

Most theoretical and even numerical studies of the intracellular transport dealt 
with a point like tracer as studied, for instance, in reference \cite{ghos15}.
In that case, one would expect that smaller crowders impede the tracer diffusion
less, as the tracer 
can always pass through even small gaps between the crowders. In turn, the point 
tracer would have to navigate around larger crowders and thus be affected more 
severely. As we showed here the opposite effect occurs
for a tracer of a finite size comparable to the size of the crowders.
In this case many small crowders significantly hamper the spreading of the tracer
as compared to larger crowders at the same crowding fraction $\phi$. 
Moreover, the van Hove correlation function acquires an exponential shape at 
higher crowding fractions and small crowders, whereas the van 
Hove correlation function is approximately Gaussian and quite insensitive 
to the crowding fraction for large crowders.

Concurrent to this effect we observe for larger, homogeneously distributed crowders 
an extended Brownian regime of the ensemble averaged MSD, whose magnitude is almost
independent of the crowding fraction within the investigated range. The time averaged
MSD in this case is also highly reproducible. For smaller crowders, however,
anomalous diffusion effects in the ensemble averaged MSD occur and become more
severe as $\phi$ increases. Additionally, individual time averaged MSD curves
demonstrate the early immobilisation of the tracer particles in the quenched
landscape, corresponding to the crossing of a local percolation threshold for the
tracer motion. In summary, when the tracer and crowder sizes are comparable, new
dynamical features emerge as compared to conventional models with point like tracers.

Similar effects were observed for heterogeneously distributed crowders.
Due to the deterministic gradient of the crowder distribution, two scenarios of 
in-out and out-in diffusion were distinguished depending on the location in which the
tracer particle was released: either at the nucleus envelope or at the cell boundary.
The in-out case resembles in many aspects the homogeneous case: the spread of the
time averaged MSD is small, the van Hove correlation function is close to Gaussian 
and weakly depends on the crowder size, and the survival probability exhibits the 
characteristic square root decay at intermediate times.  In contrast, the out-in 
case is marked by a highly non-Gaussian diffusion of the tracer: high values of 
the non-Gaussianity parameter and a pronounced cusp in the van Hove correlation 
function near $\Delta x=0$ reveal a prevalence of small displacements. This 
is characteristic for subdiffusive processes. Concurrently, the survival probability 
exhibits a very slow decay and tends to saturate at large values reflecting
the dominating fraction of crowder configurations that block the tracer near 
the release point at the cell periphery.  As a consequence, the computed averages 
over tracer trajectories are dominated by few successful translocation events.

In summary, our study revealed several important features for the tracer motion 
in a quenched disordered landscape, and is therefore also of interest from a purely 
statistical mechanical point of view. Possible applications of our results in the
fields of biological or soft matter physics concern the diffusion of tracer
particles or globular proteins in the heterogeneously crowded cytoplasmic fluid
of surface adhering (flat) cells.

In order to focus on the effects of the finite size of the tracer particle and
heterogeneous distributions of crowders, some other biologically relevant features
of living cells were ignored.
We investigated the effect of immobile crowders. On the time scale of the
motion of many passive tracers architectural elements of the cell such as parts
of the cytoskeleton or organelles may indeed be viewed immobile. Even large lipids
or insulin granules are almost localised on such time scales \cite{jeon11,tabei}.  
However, the motion of small crowders may be relevant as the effects such as 
the complete blocking of the tracer motion will be impeded.
As shown recently the major effect of small
mobile crowders is the increase of the effective viscosity experienced by the
tracer particle \cite{shin15a}. In the present study we also neglected hydrodynamic
interactions between the tracer and the crowders \cite{kapr15}
which may affect the long time behaviour of the system. The slow, $1/r$ decay of these
hydrodynamic coupling forces implies that a diffusing particle is impacted by
crowders from a finite distance that helps avoiding to collide with the crowders. 
Altering local pathways hydrodynamic forces may thus modify the diffusion
statistics. Accounting for these interactions presents an interesting perspective.
Other extensions of the current model include polydisperse 
static or dynamic crowders \cite{bara14,shah11}, cylindrical domains mimicking 
typical bacterial shapes \cite{trov14}, specific tracer-crowders interactions 
\cite{trov14,ghos15}, three dimensional simulations, variable tracer size, 
and soft polymeric crowders \cite{shin16}.

\ack

We acknowledge funding from the Academy of Finland (Suomen Akatemia, Finland
Distinguished Professorship to RM), the Deutsche Forschungsgemeinschaft (Grant
to AGC), and the French National Research Agency (Grant ANR-13-JSV5-0006-01 to
DSG).

\end{document}